# Thermodynamic properties of pure and doped (B, N) graphene

Sarita Mann[1], Pooja Rani[2], Ranjan Kumar[1], Girija S. Dubey[3] and V. K. Jindal[1*]

[1]*Department of Physics, Panjab University Chandigarh- 160014.*

[2]*D.A.V.College Sec-10, Chandigarh-160010.*

[3] *Department of Earth & Physical Sciences, York College-CUNY, Jamaica, NY 11451*

**Abstract.** *Ab-initio* density functional perturbation theory (DFPT) has been employed to study thermodynamical properties of pure and doped graphene sheet and the results have been compared with available theoretical and experimental data. The concentration of B and N has been varied upto 50% of the carbon atoms in graphene. Phonon frequencies are essential ingredients entering into such a calculation, which have been computed by using the dynamical matrix provided by VASP software in combination with phonopy code in the harmonic approximation. This easily provides us the Helmholtz free energy and leads us to numerical estimates of various thermodynamical properties. The results for specific heat are in good agreement with various theoretical and experimental studies obtained earlier for pure graphene. Interesting new results have been reported for B and N substituted structure. It has been observed that specific heat decreases with the increase in concentration of doping while the entropy increases. Further, large doping concentrations result in unstable sheets resulting in imaginary frequencies in the transverse directions. The instability needs to be compensated by external strains and that has been assumed while carrying out Brilluoin summations. These results will be useful for calculation of thermal conductivity of doped graphene and thus feasibility of using these for device applications. Our preliminary results for thermal expansion have indicated negative thermal expansion behavior of pure graphene at low temperatures which needs investigation on doped graphene as well.

**Keywords**- Graphene, density functional theory, phonons, thermodynamic properties

## 1. INTRODUCTION

The synthesis of graphene back in 2004 [1], as two dimensional thermodynamically stable crystal, led to a lot of research interest by virtue of its unique band structure and salient features [2]. It is basically a planar sheet of sp$^2$ bonded carbon atoms arranged in honeycomb lattice structure. Its various properties like ballistic transport at room temperature, high intrinsic mobility, Quantum Hall effect, combined with mechanical strength make it a potential candidate for lot of applications [3, 4]. But its use in electronics is limited by its unique band structure, which leads to the absence of a band gap. So a lot of experimental and theoretical studies have been done to modify its electronic and optical properties. We in our previous work [5, 6, 7] have done a detailed study to engineer band gap of graphene and optical properties by substituting the carbon atoms with boron and nitrogen atoms, i.e. forming a hetero structure.

---

[*] Author with whom correspondence be made, Email: Jindal@pu.ac.in

In addition to its unique electronic properties, graphene also has very high thermal conductivity [8] leading to unique transport properties. In the past few years, various experimental [8–10] reports on heat transport in graphene have shown an ultrahigh thermal conductivity, which is important for applications, such as thermal management in electronics and improvement in thermal conductance of composite materials. Properties of phonons in graphene have recently attracted significant attention from the physics and engineering researchers. Acoustic phonons are the crucial heat carriers in graphene in the vicinity of room temperature. Therefore it is important to study its thermodynamic properties in detail, which are governed by phonons which are quanta of the lattice vibrational energy.

As mentioned phonon dispersion relations are essential ingredients of thermal conductivity and for pure graphene these have been extensively studied theoretically along with thermal conductivity in recent times [11, 12, 13]. Phonon dispersion relations of graphene have also been measured experimentally using inelastic X-ray scattering [14] as well as high resolution Electron Energy loss spectroscopy (HREELS) [15] and the theoretical estimates seem to be in reasonable agreement with experimental data. Nika et. al. [13] has shown both theoretically and experimentally that transport properties of phonons are significantly different in a quasi-2-dimensional system such as graphene compared to the basal planes in graphite or three-dimensional bulk crystals. They pointed out that unique nature of two-dimensional phonon transport translates into unusual heat conduction in graphene and related materials. They described different theoretical approaches used for phonon transport in graphene discussing contributions of the in-plane and out of-plane phonon modes, and compared with available experimental thermal conductivity data. H. Zhang et. al. [16] have performed molecular dynamics simulations to investigate phonon transport in graphene at 300 K with the Green-Kubo method. They have shown a comparison of the phonon dispersions of graphene among the results from density function theory (DFT), the original reactive empirical bond-order (REBO [17] and the optimized REBO method [18]. Therefore phonon related properties of pure graphene are reasonably well documented and studied.

It has been recently pointed out that B and N doped graphene has interesting and designable electronic and optical properties [5, 6, 7] which can be suitable for device application. The issue which remains unresolved is rapidity of heat dissipation for these doped materials. In order to address this issue it is important to estimate thermal conductivity pattern of these materials. As a first step in this direction we attempt to calculate phonon dispersion and related thermodynamic properties of doped materials.

Some studies have been found in literature, which report the effect of B (N) doping on phonon related properties of graphene. Sherajul et al. [19] have presented an extensive and systematic numerical study of the vibrational properties of B- and N-doped graphene with native vacancies using the forced vibrational method. They have computed the change of the phonon density of states for different concentration of B and N impurities or vacancies without explicitly discussing the phonon dispersion relations. Yanagisawa et al. [20] have studied the phonon dispersion in stable and metastable $BC_3$ honeycomb epitaxial sheets both experimentally and theoretically.

There has been some interesting study reported on strain dependence of some phonon modes of graphene [21] as well as of silicene [22]. Further, Jian et al. [23] have done detailed study (upto 50 %) of B and N doped graphene and discussed the stability of such structures on the basis of negative frequencies which occur in these structures. They emphasize that if appropriate strain as well as charge is applied to the structures at high doping concentration, it leads to positive frequencies. But to the best of our knowledge, there is not any systematic study, describing the pattern of changes in phonon dispersion curve because of doping and thermodynamic properties resulting from doped structure like specific heat, entropy and free energy, found in literature so far. So in the present work, we attempt to analyze the effect of boron and nitrogen substitution on stability and phonon dispersion curves of graphene sheet and calculated various thermodynamic properties of pure and doped graphene. While discussing thermodynamic properties, as pointed out by Jian et al., negative frequencies do occur indicating the instability of heavily doped graphene sheet in the transverse direction. Assumption of a strain in the transverse direction can be used to provide stability and remove negative frequencies.

The paper is structured as follows. After summarizing the theoretical formulation and computational details in sections two, we discuss the results in section three. The results are summarized and concluded in section 4.

## 2. THEORY AND COMPUTATIONAL DETAILS

To calculate the phonon dispersion and thermodynamic properties we have made use of VASP (Vienna *ab- initio* Simulation Package) [24-25] code based on density functional theory (DFT) in interface with phonopy [26] code. In density-functional theory, the ground state electronic density and wave functions of a crystal are found by solving self-consistently a set of one-electron equations. These procedure and equations are well described in literature[27] .The Kohn-Sham equation is solved self consistently to find the total energy of the system. The VASP program is run to accumulate data of forces at displaced structures which will be used in subsequent sections.

### 2.1) Dynamical Matrix

In order to obtain phonon frequencies we need to obtain the interatomic force constants $\Phi_{\alpha\beta}$ [28-30], i.e.

$$\Phi_{\alpha\beta}(l\kappa,l^{'}\kappa') = \frac{\partial^2 \phi}{\partial r_\alpha(l\kappa) \partial r_\beta(l^{'}\kappa')} \qquad (1)$$

Where $l\kappa$ are positions of any atom w. r. t other atom at $l'\kappa'$. $\alpha$ and $\beta$ are Cartesian indices, **r** is position vector of atom . In order to obtain the force constants, we further express them in terms of forces

$$\Phi_{\alpha\beta}(l\kappa, l'\kappa') = -\frac{F_\beta(l'\kappa'; \Delta r_\alpha(l\kappa)) - F_\beta(l'\kappa')}{\Delta r_\alpha(l\kappa)} \qquad (2)$$

where $F_{\beta(l\kappa)}$ is the force on the $l\kappa$ atom in $\beta$ direction and $\Delta r_\alpha$ is finite displacement. Since the VASP program gives us the forces, therefore equation (2) helps us to obtain force constants $\Phi_{\alpha\beta}$ and the dynamical matrix [26]. Diagonalizing the dynamical matrix gives us phonon frequencies $\omega_{\mathbf{q}i}$, for wave vector **q** and branch *i*.

## 2.2) Thermodynamical properties

In a harmonic crystal, the structure does not depend on temperature. The expressions for heat capacity per unit cell at constant volume and entropy (S) [26] are obtained by summing over all the phonon branches.

$$C_v(T) = k_B \sum_{\mathbf{q}i} \left(\frac{\hbar\omega_{\mathbf{q}i}}{K_B T}\right)^2 \frac{\exp(\hbar\omega_{\mathbf{q}i}/k_B T)}{(\exp(\hbar\omega_{\mathbf{q}i}/k_B T) - 1)^2} \qquad (3)$$

$$S = -\frac{\partial F}{\partial T} = \frac{1}{2T}\sum_{\mathbf{q}i} \hbar\omega_{\mathbf{q}i} \coth\left(\frac{\hbar\omega_{\mathbf{q}i}}{2K_B T}\right) - k_B \sum_{\mathbf{q},i} \ln\left[2\sinh\left(\frac{\hbar\omega_{\mathbf{q}i}}{2K_B T}\right)\right] \qquad (4)$$

The expression for free energy term [29] used here for calculating free energy as given below

$$F(T) = \phi(V_0) + k_\mathbf{B} T \ln\left(2\sinh\frac{\hbar\omega_{\mathbf{q}i}}{2K_B T}\right) \qquad (5)$$

where $\phi(V_0)$ is the ground state energy at the relaxed volume $V_0$.

## 2.3) Computational parameters

The graphene sheet consisting of 32 atoms has been used for simulations and sheets are separated by larger than 9.8Å along perpendicular direction to avoid interlayer interactions. We adopted the Perdew-Burke-Ernzerhof (PBE) [31] exchange correlation (XC) functional of generalized gradient approximation (GGA) in our calculations. The plane wave cut-off energy was set to 750eV for graphene. The Monkhorst-pack scheme [32] is used for sampling Brillouin zone. The structure is fully relaxed with Gamma centered 7 x 7 x 1 k-mesh. The partial occupancies were treated using the tetrahedron methodology with Blöchl corrections [33]. For geometry optimizations, all the internal coordinates were relaxed until the Hellmann-Feynman forces were less than 0.005 Å.

The phonopy software is used in combination with VASP to produce phonon frequencies. To obtain the force constants (FCs) in phonopy, we manually displace the atoms in the center cell by $\Delta x^\alpha$ ($\alpha$ denotes the three directions), and then calculate the IFCs of the perturbed structure.

## 3. RESULTS AND DISCUSSION

### 3.1 Phonon Dispersion

The investigation of lattice vibrational modes is the initial step to study the thermodynamic properties of a material. The simplest approximation is harmonic approximation, where all the oscillators in a solid are harmonic and have fixed frequencies. Using eq (3) and diagonalizing the dynamical matrix we obtain phonon dispersion curve. The dispersion is basically the relationship between the phonon frequency and the phonon wave vector **q**. In this section phonon dispersion results for pure graphene have been presented and compared with available theoretical and experimental data. After the successful reproduction of results for pure graphene, the phonon dispersion curves for doped graphene have also been obtained.

### 3.1.1 Graphene unit cell (2-atoms)

Graphene is a hexagonal 2D sheet of carbon atoms arranged in a honeycomb lattice with two atoms in the unit cell as shown in Fig.1 where **a₁** and **a₂** are lattice vectors such that

$|a_1|= |a_2|=$ lattice constant $=a_0=2*b\cos 30^0=\sqrt{3}b$   (b=bond length)

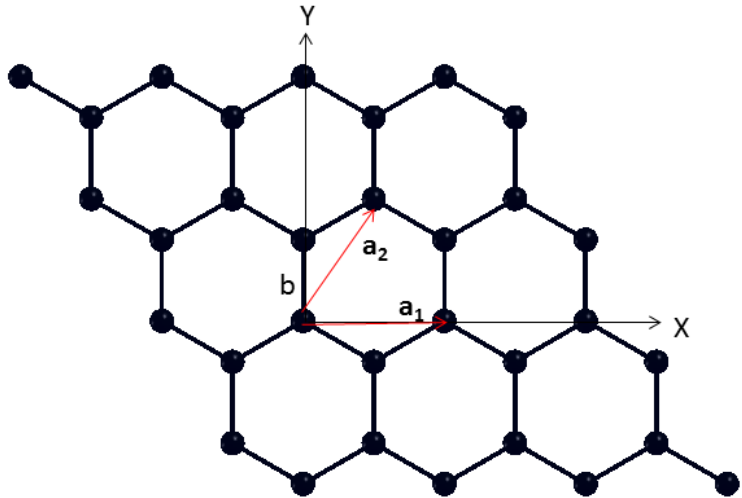

FIGURE 1. Schematic representation of 2D Lattice structure of Graphene with **a₁** and **a₂** represent the lattice vectors of unit cell and b shows the bond length at the edge of hexagon

In a lattice with a basis of two atoms in the primitive cell, there will be four allowed frequencies of wave, two upper branches known as the longitudinal and transverse optical branches abbreviated as LO and TO, and two lower branches called longitudinal and transverse acoustical branches abbreviated as LA and TA respectively. Acoustic phonons have frequencies that become smaller at long wavelengths, and correspond to sound waves in the lattice. We artificially managed a distance of approximately 10 Å between two graphene sheets and displaced the atoms in z-direction also which generates out of plane transverse acoustic (ZA) and optical (ZO) varieties respectively. Thus we get 6 phonon branches in all for a unit cell of graphene. We now discuss in the following subsections phonon dispersion of pure and doped graphene.

**3.1.1.1 Pure graphene**

In this section, using the theoretical approach described above, we compute the phonon dispersion in single layer graphene which is shown in Fig.2. The key feature of phonon dispersion in graphene is the existence of ZA mode also termed as flexural mode [34] which is the lowest frequency mode and easiest to be excited. The origin of this mode is only due to the surface interactions. The behavior of ZA mode is quadratic in the vicinity of gamma point. There are other important key features of such modes which are derivable from longitudinal modes and exist only for low dimensional systems [35]. In fact both ZA and ZO originate from longitudinal motion but ZA is more important because it is the lowest frequency mode.

The phonon dispersion curves obtained above are in good agreement with phonon dispersions calculated using VASP by Zhang et al. [16]. Zhang et al. have plotted the phonon dispersion curve using VASP and REBO potential for pure graphene. Our results are also similar to [36] where the phonon dispersion is calculated by constructing a dynamical matrix using the force constants derived from the second-generation reactive empirical bond order potential by Brenner and co-workers. Finally we also compared our results with experimental data obtained by Mohr et al. [14]. They have studied phonon dispersion by Raman 2D-band splitting in graphene and Yanagisawa et al. [15] have used high resolution electron energy loss spectrometer technique for calculating the phonon dispersion in graphene. Result of experimental technique [15] matches with our calculation as shown in Fig.2.

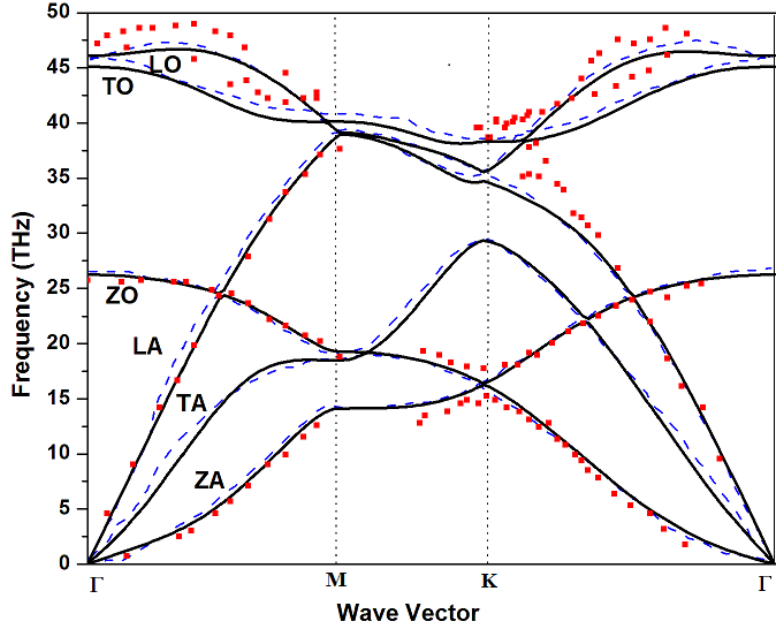

FIGURE 2. Phonon dispersion curve calculated for pure graphene (solid lines) as compared to that in Ref. [16] (Theory) (dashed lines) and Ref. [15] (symbols) (Experimental). LA(LO), TA(TO) and ZA(ZO) are longitudinal, transversal and out-of-plane acoustical (optical) branches respectively.

We also calculated the phonon frequencies for different approximations of exchange correlations. The data is presented along with values provided by Mounet et al. (theory) [37] and Yanagisawa et al. [15] (Experimental) in table 1. All the acoustic modes vanish at $\Gamma$ point.

**Table 1.** Calculated phonon frequencies for the $\Gamma$, M and K points

| High Symmetry Point | Frequency (THz) | | | | |
|---|---|---|---|---|---|
| | LDA | GGA-PBE | GGA-91 | Ref. [37] (GGA-PBE) | Ref.[15] (Experimental) |
| $\Gamma_{ZO}$ | 26.90 | 26.13 | 26.23 | 26.41 | 26.02 |
| $\Gamma_{LO}$ | 45.99 | 46.45 | 46.05 | 46.58 | 47.36 |
| $\Gamma_{TO}$ | 46.11 | 46.57 | 47.02 | 46.58 | 47.36 |
| $M_{ZA}$ | 14.44 | 14.00 | 13.97 | 14.12 | 13.52 |
| $M_{TA}$ | 18.67 | 18.61 | 18.52 | 18.76 | |
| $M_{LA}$ | 39.33 | 39.54 | 39.48 | 39.81 | 39.81 |
| $M_{ZO}$ | 19.47 | 18.88 | 19.11 | 19.03 | 19.03 |
| $M_{LO}$ | 39.73 | 40.25 | 40.01 | 40.17 | 39.66 |
| $M_{TO}$ | 40.78 | 41.36 | 41.17 | 41.67 | 41.67 |
| $K_{ZA}$ | 16.44 | 15.90 | 16.10 | 16.03 | 15.49 |
| $K_{ZO}$ | 16.73 | 16.20 | 16.18 | 16.03 | 17.62 |
| $K_{TA}$ | 29.49 | 29.47 | 29.48 | 29.88 | |
| $K_{LA}$ | 35.33 | 35.64 | 35.41 | 36.36 | 36.36 |
| $K_{LO}$ | 35.91 | 36.23 | 36.27 | 36.36 | 38.61 |
| $K_{TO}$ | 38.90 | 39.59 | 39.47 | 38.61 | 38.61 |

The comparison of phonon frequencies for different exchange correlation shows no significant variation. Therefore this establishes that our procedure is appropriate for computation of phonon frequencies and therefore we extend our computation to phonon dispersion of doped graphene.

### 3.1.1.2 Doped graphene

We started with a doping of 50% in pure graphene which is obtained by replacing one atom of carbon with boron (nitrogen) in a unit cell of graphene and then repeating the unit cell along xy direction to prepare a 32 atom $h$-CB/$h$-CN (hexagonal) sheet. We analyzed the ordered $h$-CB and $h$-CN in its neutral state and their phonon dispersion curves, obtained by using similar the technique as for pure graphene, are presented in Fig. 3. Fig. 3(a) shows the phonon dispersion of relaxed structure with equilibrium lattice constant $a_0$ as 2.698Å. Fig. 3(b) shows the phonon dispersion of relaxed $h$-CN structure with equilibrium lattice constant $a_0$ as 2.397 Å. Phonon dispersion curves for $h$-CB shows that the lower ZA branch has imaginary frequencies (following conventional notation, we plot imaginary frequency as negative values) indicating that the structure is unstable. The phonon dispersion curve of *h*-CN shows two branches having imaginary frequency modes. The lower soft branch corresponds to out-of-plane acoustic (ZA) mode, and the upper branch is out-of-plane optic (ZO) mode.

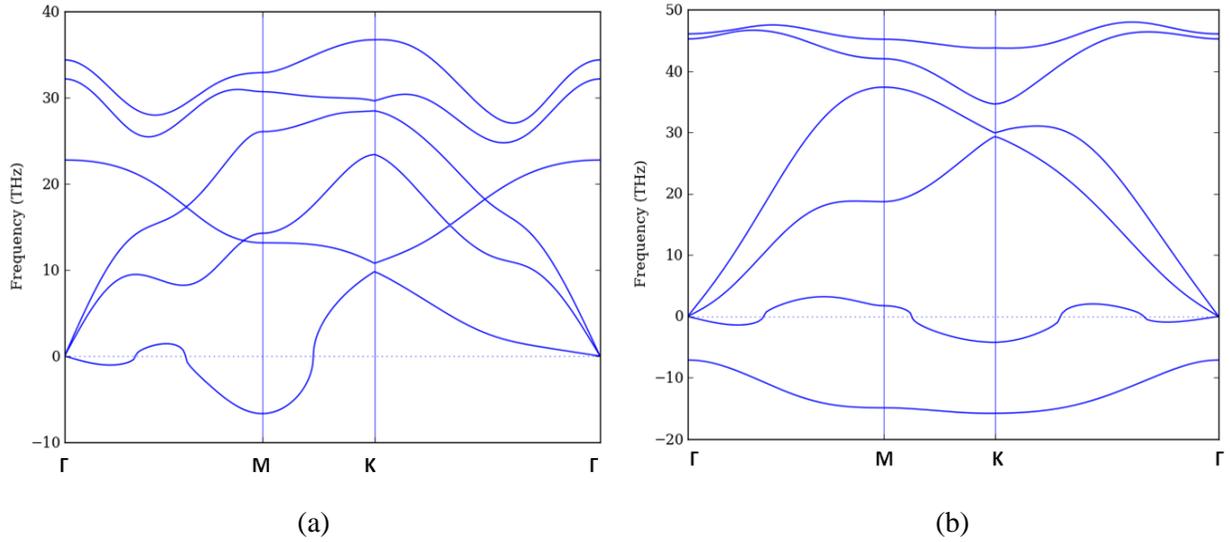

(a)            (b)

FIGURE 3. Phonon Dispersion curves for 50% (a) B-doped graphene (b) N-doped graphene

The imaginary frequency modes indicate that the neutral *h*-CB and *h*-CN are dynamically unstable, that is why it is difficult to synthesize these structures. Although we relaxed our structure but the nature of phonons confirms that it is not a stable structure. We therefore conclude that 50% doped structure is unstable under normal conditions.

As discussed earlier Zhou et al. [23] have shown that these structures can be stabilized by applying appropriate strains alongwith doping with opposite charge. They modify the bond length to introduce strains. Therefore in the following section we assume that such doped structures are possible to be stabilized by applying strain alongwith the charges.

### 3.1.2 Graphene unit cell (8-atoms)

In order to vary the concentration of doping, it is necessary to take bigger unit cell. We take unit cell comprising of 8 atoms which allows us to vary the concentration of dopants from 1/8 i.e.12.5% onwards upto 4/8 i.e. 50%. The possible configurations considered by us for the present calculations are shown in Fig.4. Each structure is initially relaxed and lattice constant is obtained for minimum energy configuration. The electronic properties of such structures have already been investigated by our group [5].

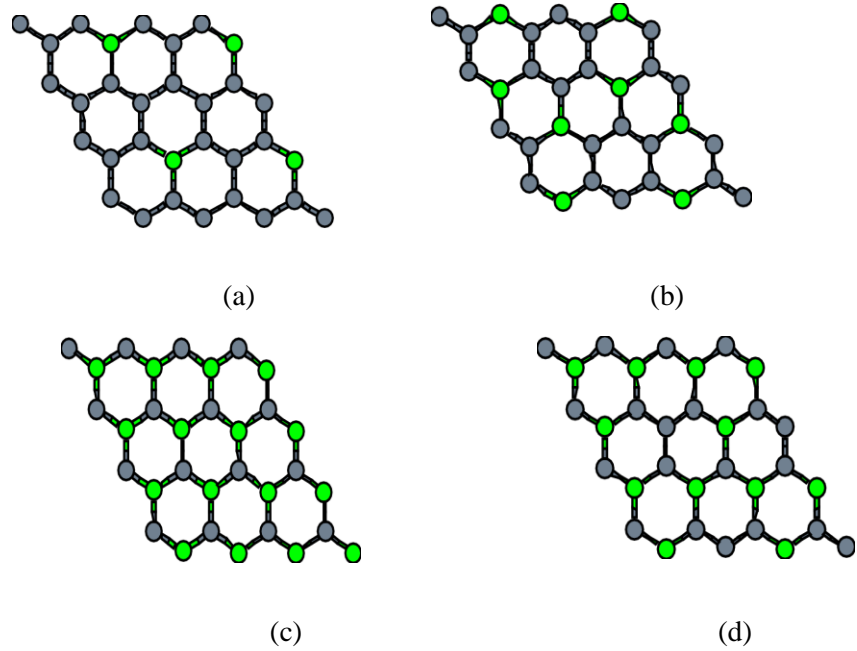

(a)　　　　　　　　　　　　(b)

(c)　　　　　　　　　　　　(d)

FIGURE 4. Configurations of (a) 12.5% (b) 25% doped, (c) 37.5% and (d) 50% doped (B, N) relaxed structures

### 3.1.2.1 Pure graphene

The phonon dispersion of pure graphene for a graphene unit cell of 8 atoms has been shown below in Fig. 5, which gives rise to 3N phonon branches i.e. 24. This dispersion is similar to the one calculated above for a unit cell (2-atoms) of graphene. However there are artificially generated (degenerate) branches which can be correlated with six branches of Fig. 2.

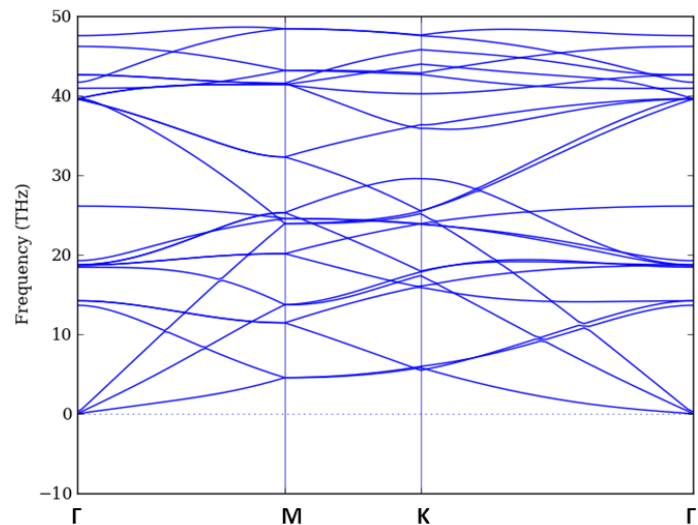

FIGURE 5. Phonon dispersion curves of pure graphene for 8-atoms unit cell

### 3.1.2.2 Doped graphene

By varying the concentrations as described in Fig.4 we obtained the graphene structures with different doping concentrations. We optimized these structures for minimum forces but they do not give a correct picture of stability in z-direction, this will be manifested once we discuss the results of phonon frequencies. The lattice constant for an 8-atom unit cell becomes double the original lattice constant. i.e. $a_0' = 2a_0$. In case of B-doping, the lattice constant $a_0'$ keeps on increasing from (2x2.46) 4.92 Å for pure graphene to 5.34 Å for 50% doped structure. Inversely the lattice constant $a_0'$ for N-doped structure decreases from 4.92 Å to 4.79 Å for 50% doped structure. The structure with 12.5% and 25% doping of B and N atoms separately in graphene sheet are stable with no imaginary frequencies and it becomes slightly unstable as we increase the percentage of doped atoms to 37.5% and highly unstable for 50% doped structure as shown in Fig.3 earlier. The variation of lattice constant with doping is given in table 2.

**Table 2.** Variation of Lattice constant with doping Concentration

| Doping Conc. | | 12.5% | 25% | 37.5% | 50% |
|---|---|---|---|---|---|
| Lattice Constant ($a_0'$) | B-Doped | 5.04 | 5.17 | 5.25 | 5.34 |
| | N-Doped | 4.90 | 4.86 | 4.82 | 4.79 |

The study of phonon dispersion of B doped graphene for these doping concentrations is shown in Fig. 6.

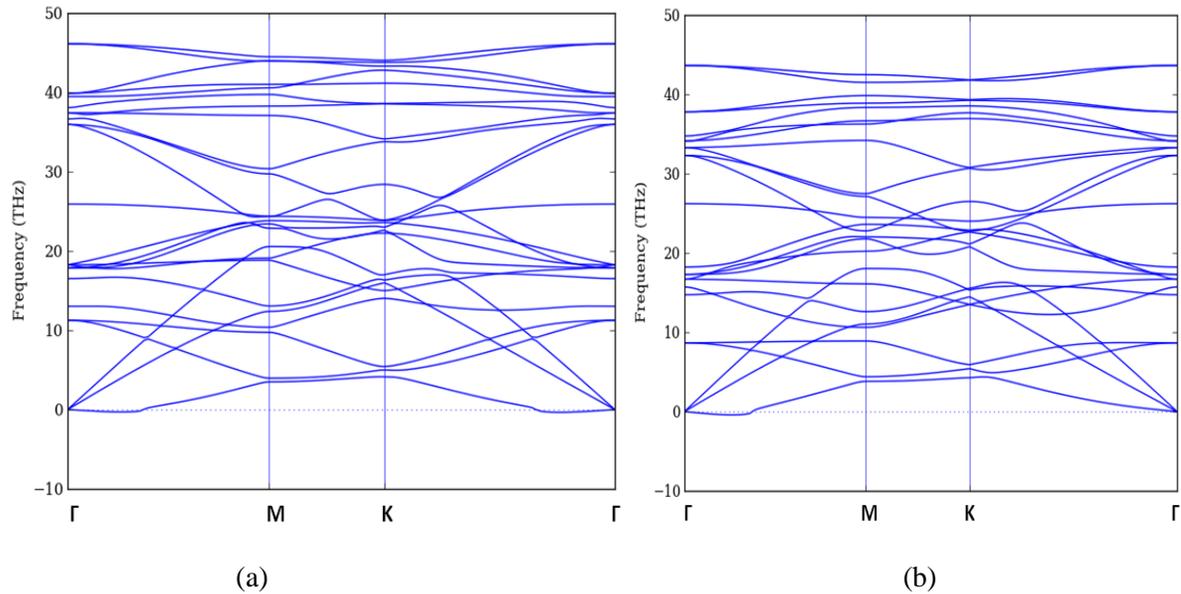

(a)  (b)

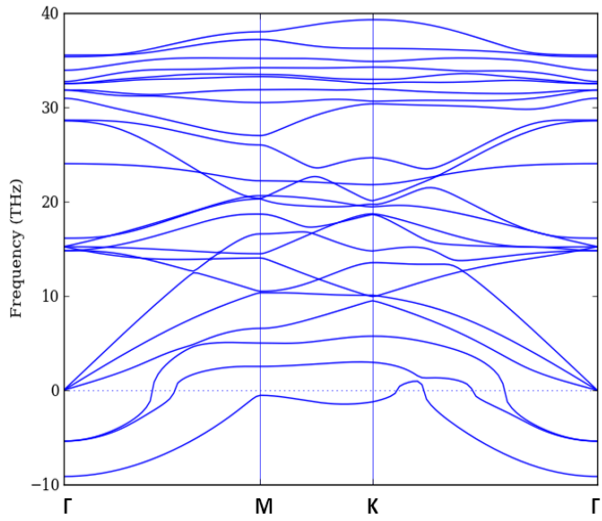
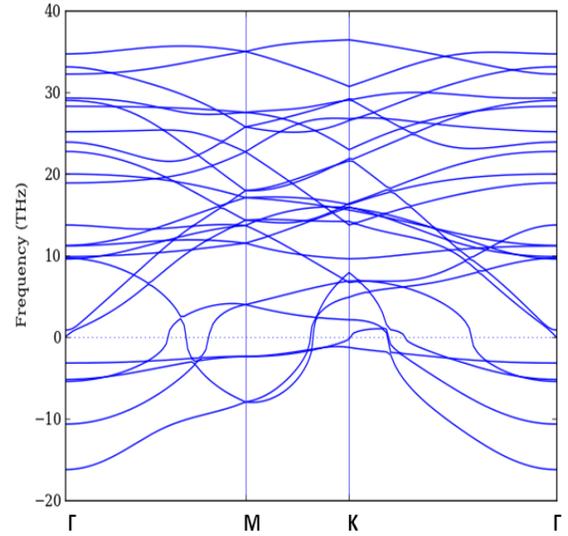

(c)                                            (d)

FIGURE 6. Phonon dispersion curve of B doped graphene (a) 12.5%  (b) 25%  (c) 37.5% and (d) 50% respectively.

Similarly, the phonon dispersion curves for N doped graphene for various doping concentrations have been obtained and shown in Fig. 7.

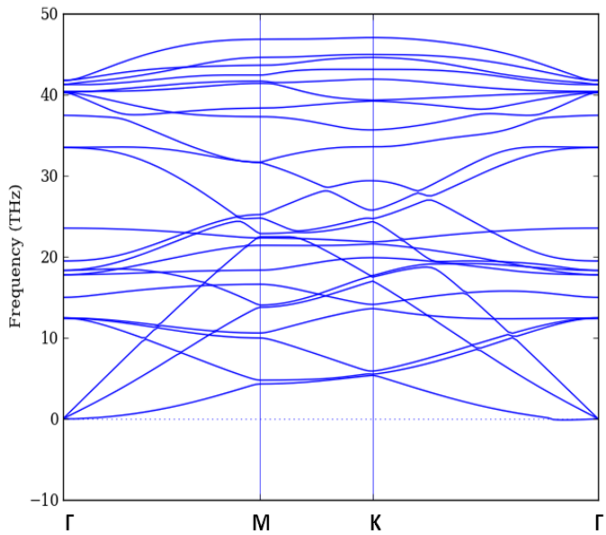
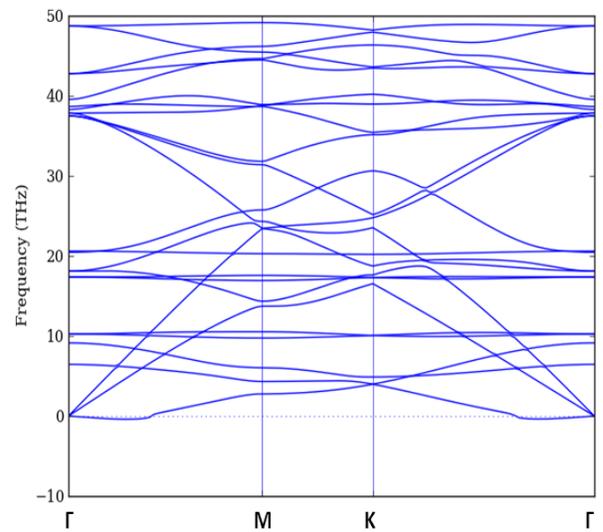

(a)                                            (b)

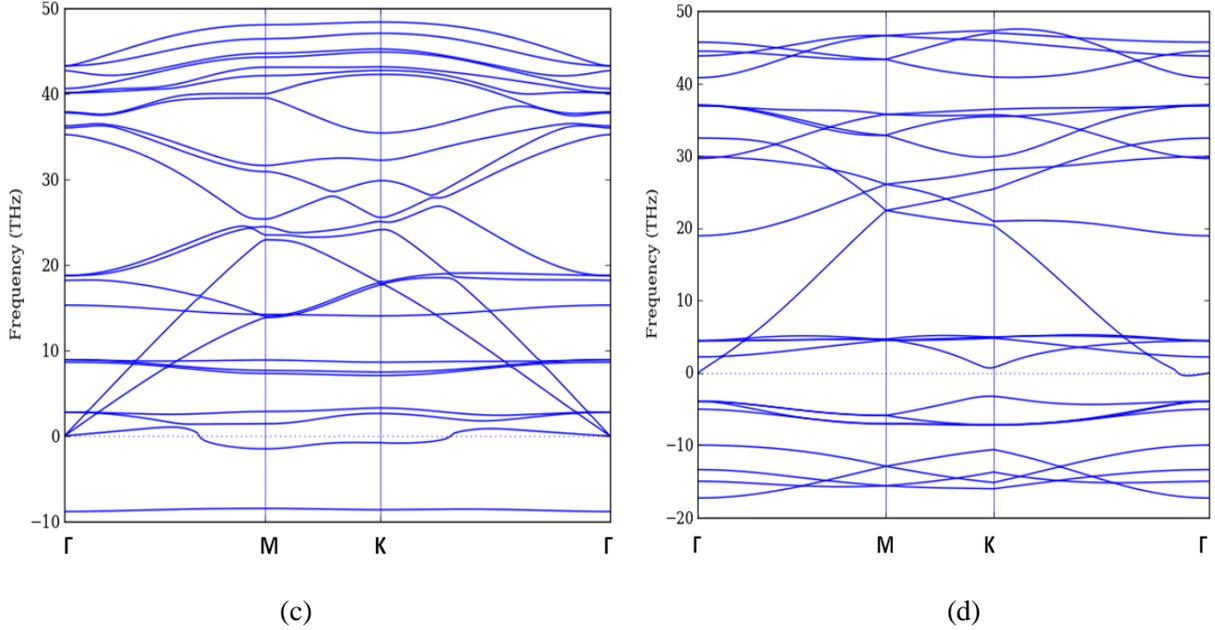

(c)                                           (d)

FIGURE 7 . Phonon dispersion curve of (a) 12.5% (b) 25% (c) 37.5 and (d) 50% N-doped graphene respectively.

We observe that for the case of 12.5% doping (Fig. 6(a) (B-doping) and in Fig. 7(a) (N-doping), all the frequencies are positive but they no longer are degenerate branches as in pure graphene case in Fig. 5. Thus we have separate branches with even minimal of doping.

When we move to 25% doping in Fig. 6(b) (B-doping) and in Fig. 7(b) (N-doping), the phonon dispersion shows a softening of acoustic and optical branches. The frequency of lowest ZA branch is decreased in both the cases.

For the higher doping concentrations i.e. 37.5% and 50% as shown in Fig. 6(c & d) (B-doping) and in Fig. 7(c & d) (N-doping), we get imaginary frequencies which are more in number in 50% doped case. There is a noticeable feature that the maximum phonon frequency range largely reduces in case of B doped strucure from 48 THz to 35 THz. Although there is decrease in maximum frequency in N doped strucures also but there is no prominent change. Therefore as far as for these concentrations are concerned, our structure is unstable. Further, the modes which became imaginary are ZA and ZO branches only.

**3.2 Phonon DOS**

The DOS curves are plotted in Fig. 8 for doping concentrations from 12.5% to 50% as compared to pure graphene. In obtaining DOS and various thermodynamic properties, we assumed that by applying proper value of strain and charging as reported by the work done by Zhou et al. [23], we get the structures with nearly zero frequencies of ZA and ZO branches in case of 37.5% and 50% doped structures which have imaginary frequencies.

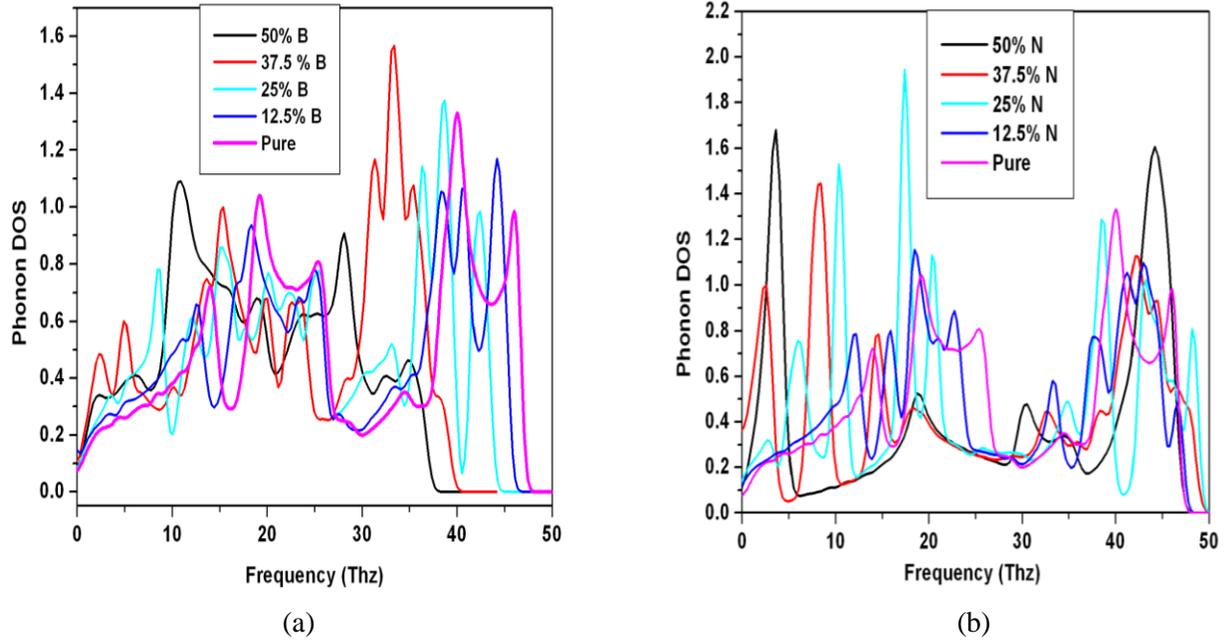

(a)                                        (b)

FIGURE 8. Phonon density of states for (a) B-doped and (b) N-doped structures with varying doping concentration along with pure graphene.

The comparison of phonon density of states clearly indicates the difference from pure graphene case. As the B and N doping is increased, the peaks corresponding to pure graphene are shifted towards lower frequency and the intensity of peaks also increased. The DOS also indicates formation of some new peaks in addition to pure one as reported earlier in Ref. [19].

### 3.3 Lattice Thermodynamical Properties

Thereafter we computed thermodynamical properties of pure graphene and compared with doped graphene. The specific heat of pure graphene as calculated by us has been compared with experimental data [39] as well as with earlier calculations[37,38] and presented in Fig. 9. It is observed that there is good agreement with the experimental data. The specific heat, entropy and free energy are calculated using equations (6-8) and have been presented in Fig 10 and 11. For the case of high doped graphene where the imaginary frequencies are encountered, we take them to be zero to calculate thermodynamical properties. This assumption is justifiable in view of an earlier proposal [23] of disappearance of negative frequencies under appropriate strain and charging. We have assumed that these heavily doped graphene are under appropriate strain and charging which turns the negative frequencies of these structures to nearly zero frequencies.

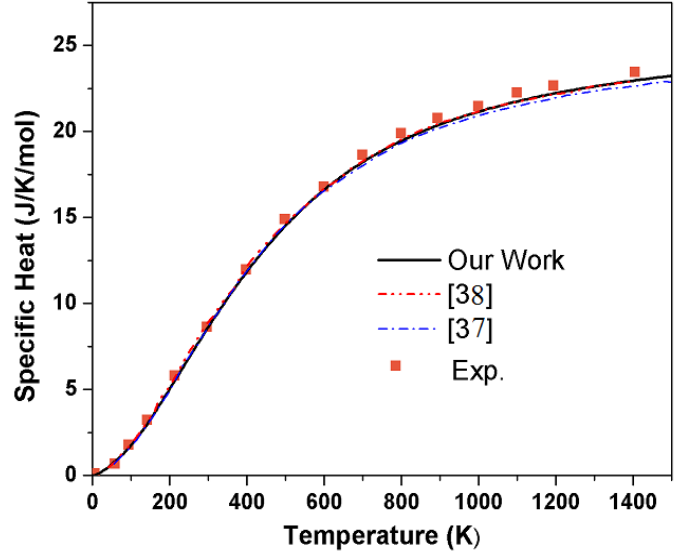

Fig.9. Specific heat of pure graphene at various temperatures compared with experimental [39] and other theoretical calculations [37, 38].

It is easily observed from the graph as shown in Figs. 10(a) and Fig. 11 (a) that specific heat for pure graphene as well as doped graphene reaches a constant value 21 J/K/mol which is close to its Dulong-Petit value at high temperature (1000 K) [38, 39], however near room temperature (300 K) it is around 7 J/K/mol (0.6 J/g/K) for case of pure graphene. We notice that entropy and free energy show systematic pattern in their dependence in case of B doped graphene. The behavior of all the thermodynamic properties for the case of N doped graphene show some deviations at certain temperatures as shown in Figs. 11(a, b, c). Due to soft modes in the transverse direction, there is a tendency for specific heat to increase at low temperature and decrease at high temperatures for highly doped configurations. Unfortunately, there is no experimental or other theoretical data on the lattice dynamical properties of graphene for the sake of comparison.

The specific heat, entropy and free energy values of 12.5% B and N doped graphene are close to that for pure graphene. The specific heat of doped graphene increases with doping at low temperatures for all doping concentrations. The increase in specific heat can be accounted to the higher phonon density of states at low frequencies as shown in Fig. 10. The behavior of specific heat for doped graphene decreases to a constant value above cut off (Debye) temperature. But the specific heat for 50% B and N doped graphene structures is strikingly becoming constant at a much lower value at high temperatures. 50% doping structure shows peculiar behavior in N doped structure. This may be due to highly suppressed transverse modes.

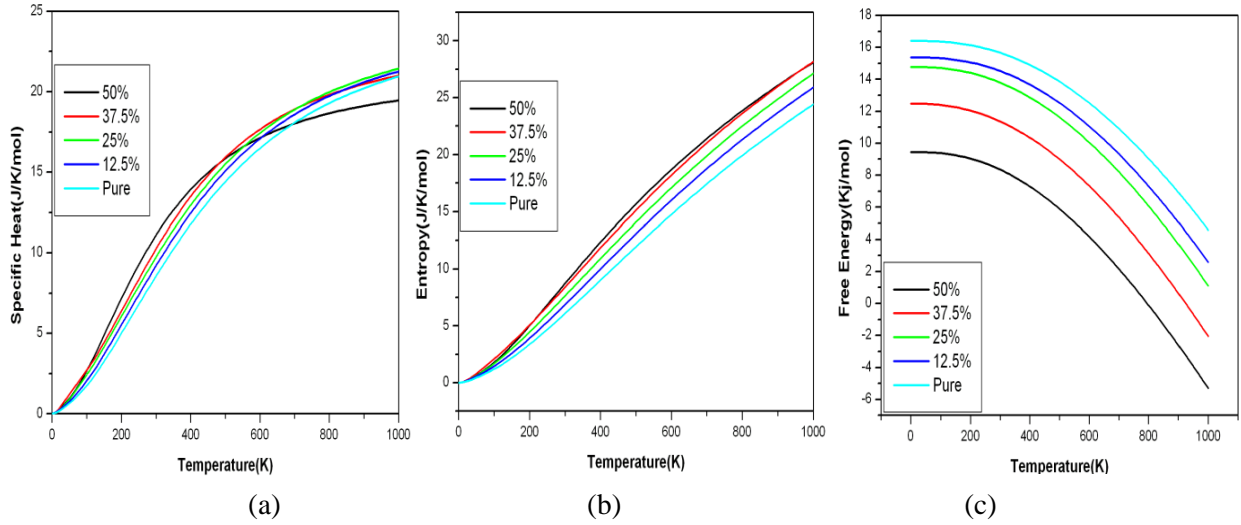

FIGURE. 10 Variation of Specific Heat, Entropy and Free energy with temperature for B doped graphene with increase in doping concentration

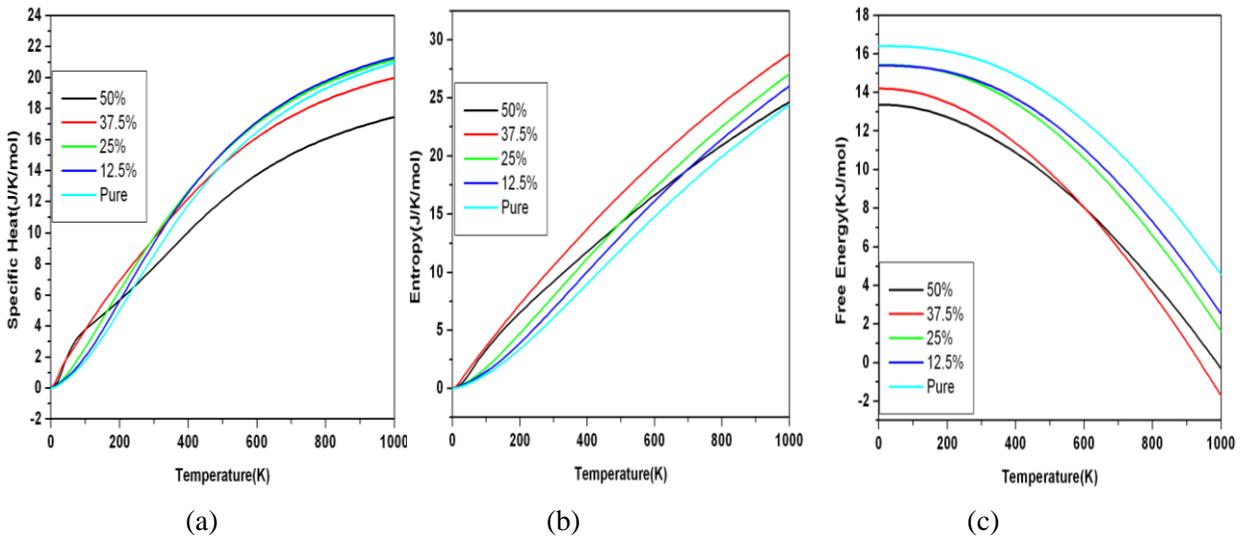

FIGURE. 11 Variation of Specific Heat, Entropy and Free energy with temperature for N doped graphene with increase in doping concentration

The entropy is increasing with increase in doping concentration for both B and N doped structures. As the B doping concentration is increased to 50%, there is large increase in entropy as compared with pure graphene case. But there is a significant decrease in entropy of 50% N doped graphene at higher temperatures approaching the pure graphene case, indicating a favorable condition for formation of the *h-CN* sheet at higher temperatures, whereas there is no such decrease in value of entropy in case of B-doped graphene. This is probably due to similar size of C and N, which favors the formation of stabilized h-CN sheet at high temperatures which is indicated by the increase in entropy.

The variation of free energy with temperature for various doping concentrations (Figs. (10 (c) and 11 (c)) shows a constant decrease with increase in B(N) doping concentrations. Unlike, in case of N doping, the behavior is similar for all doping concentrations except 50%. The 50% doped curve for free energy is lower at low temperature values and cuts the 37.5% curve at around 600K and becomes higher at high temperatures. Thus free energy curve of 50% N doped structure does not decrease as fast as it should happen. Overall we observe a systematic behavior (increase or decrease) for all the thermodynamic quantities with increasing the doping concentration starting from (12.5%). However, this trend changes unusually at 50% doping concentration, which is more prominent in case of N-doped structures.

We also performed preliminary calculations for Grüneisen parameter defined as

$$\gamma_{qi} = -\frac{V_0}{\omega_{qi}}\frac{\partial \omega_{qi}}{\partial V} \quad (6)$$

The Grüneisen parameters are a measure of anharmonicity of the potential and are useful parameters for determining thermal expansion and implicit anharmonic properties [30].

As shown in Fig. 12 where we present an estimate of Grüneisen parameters of various modes and thus calculated thermal expansion at various temperatures compared with available theoretical and experimental data. Although ZA mode is likely to result in negative value of Grüneisen parameter [35], but such high negative value of Grüneisen parameter for ZA mode is unusual and requires further investigation. Although our results reproduce the results for thermal expansion qualitatively, but there is significant variation in experimental and other theoretical data. There is negative thermal expansion reported in the low temperature region but there are significant variations. The discrepancies between our study and other theoretical [37, 41] and experimental [40] studies also accordingly needs to be fully understood. Once this work is investigated, it will be extended to the doped graphene.

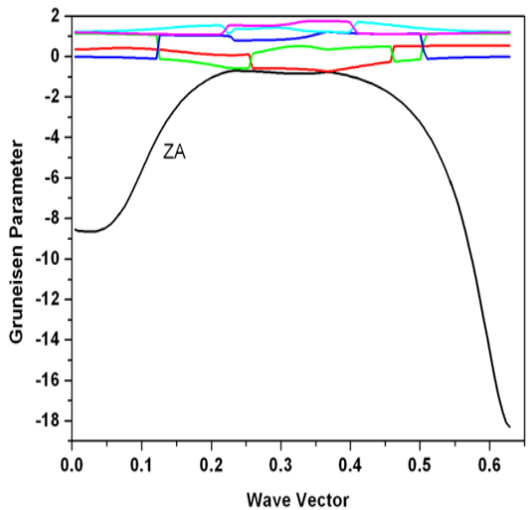
(a)

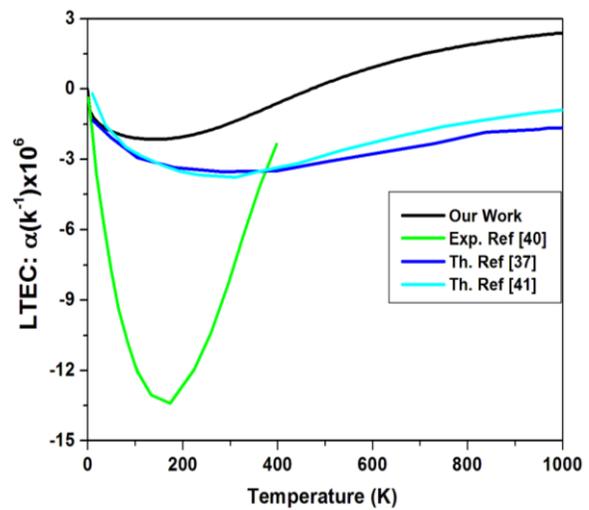
(b)

Fig.12. Grüneisen parameters of various modes in pure graphene at various q values (a) and resulting thermal expansion at various temperatures (b) including other results.

## 4. SUMMARY AND CONCLUSIONS

We have presented a detailed first-principles study of pure and B and N doped graphene at the GGA-PBE level under harmonic approximation to derive the finite temperature behavior of several thermodynamic quantities. All our results for pure graphene sheet are in very good agreement with previous theoretical and experimental data. The 2-D graphene structure which was optimized under static conditions does not necessarily result in stable 2-D configuration, especially on large doping concentrations. This is evidenced by the results of transverse phonon frequencies which begin to appear as imaginary above critical concentrations of B as well as N. In order to retain these structures as stable which return positive frequencies in the transverse directions as well, we follow Zhou et al.[23] and assume a strained and charged 2-D lattice compensates for this instability. Therefore the role of imaginary frequencies can be overlooked as has been done here. Various thermodynamic properties of B and N doped graphene at different concentrations have been compared with the pure graphene. In general, the heat capacity and free energy of the structures decreases and entropy increases with increasing doping concentration as compared to pure graphene while the entropy( heat capacity, free energy) of the 50% doped structure is dramatically low (high) which is possibly due to the higher symmetry and hence stability of this structure. The heat capacity of doped graphene is higher at low temperatures and decreases to a constant value at high temperature limit as compared to pure graphene. The higher specific heat of doped graphene can be explained on the basis of higher value of density of states at low frequencies. The increasing value of entropy fairly explains the increasing disorder and decreasing stability with doping. The decrease in entropy and increase in heat capacity of the 50% N doped structure is due to symmetry and similar sizes of carbon and nitrogen. Our results suffer from not using appropriate strained lattices under high doping which stabilizes the 2-D structure in the transverse direction and we intend to examine this.

## ACKNOWLEDGEMENTS

We express our gratitude to VASP and phonopy team for providing the code, the HPC facilities at IUAC (New Delhi) and the departmental computing facilities at Department of Physics, Panjab University, Chandigarh.